\documentclass[preprint,showpacs]{revtex4}
\usepackage{graphicx,amsmath,amssymb,amsfonts,latexsym,color,dcolumn,bm}

\begin{document}
\title{Casimir-Polder interatomic potential between two atoms at finite temperature
and in the presence of boundary conditions}

\author{R. Passante}
\email{roberto.passante@fisica.unipa.it}
\author{S. Spagnolo} \affiliation{Dipartimento di Scienze Fisiche ed
Astronomiche dell'Universit\`{a} degli Studi di Palermo and CNISM,
Via Archirafi 36, I-90123 Palermo, Italy}

\pacs{12.20.Ds, 42.50.Ct}

\def\br{{\bf r}}
\def\bk{{\bf k}}
\def\ekj{\hat{e}_{\bk j}}
\def\akj{a_{\bk j}}
\def\akjd{a_{\bk j}^\dagger}
\def\FlmR{(\nabla^2\delta_{lm}-\nabla_l\nabla_m)^R}
\def\Rbar{{\bar{R}}}

\def\bE{{\bf E}}
\def\bR{{\bf R}}
\def\Fln{F_{\ell n}}
\def\Flm{F_{\ell m}}
\def\Fmn{F_{mn}}
\def\bgsC{\mid vac, \downarrow_C\rangle}

\date{\today}

\begin{abstract}
We evaluate the Casimir-Polder potential between two atoms in the
presence of an infinite perfectly conducting plate and at nonzero
temperature. In order to calculate the potential, we use a method
based on equal-time spatial correlations of the electric field,
already used to evaluate the effect of boundary conditions on
interatomic potentials. This method gives also a transparent
physical picture of the role of a finite temperature and boundary
conditions on the Casimir-Polder potential. We obtain an
analytical expression of the potential both in the near and far
zones, and consider several limiting cases of interest, according
to the values of the parameters involved, such as atom-atom
distance, atoms-wall distance and temperature.
\end{abstract}

\maketitle

\section{Introduction}
Casimir-Polder forces are long-range interactions between neutral
atoms or molecules due to their common interaction with the
electromagnetic radiation field. In the case of two atoms in the
vacuum (zero temperature) the Casimir-Polder potential energy
behaves as $R^{-6}$ for interatomic distances smaller than typical
atomic transition wavelenghts from the ground state (near zone)
and as $R^{-7}$ for larger distances (far zone) \cite{CT98}. In
the near zone the potential energy coincides with the well-known
van der Waals interaction, but in the far zone it decreases more
rapidly due to retardation effects. Analogous interactions exist
between an atom and a neutral conducting wall and between two
conducting or dielectric walls (the so-called Casimir effect)
\cite{Milonni94}. These interactions are usually considered as a
manifestation of the quantum nature of the electromagnetic
radiation field and related to the zero-point energy. Although
Casimir-Polder and Casimir energies are very small, the Casimir
force between macroscopic objects has been measured with
remarkable precision (for a review, see \cite{Lamoreaux05}).
Relevance of Casimir forces to nano- and micro-devices has been
also shown \cite{CAKBC01,ILMC05}. Also the atom-wall
Casimir-Polder force has been recently measured with precision,
both in the near and in the far zone
\cite{Sukenik1,Sukenik2,DruzhDeKie,HOMC05,OWAPSC07}. The atom-atom
van der Waals/Casimir-Polder energy is still weaker, but
experimental indirect evidences of them exist since a long time,
in agreement with theoretical predictions \cite{VO03}. Direct
measurements of the retarded van der Waals attraction in
mesoscopic systems have been also obtained \cite{CT78,BP99}. In
order to obtain direct high-precision evidence of the atom-atom
force, it can be relevant to evaluate them in realistic situations
to be compared with actual laboratory situations, for example by
taking into account temperature effects and/or the presence of
boundary conditions, as well as to envisage situations where the
intensity of these forces could be increased. In a previous paper
we have calculated the atom-atom Casimir-Polder interaction energy
when the two atoms are placed in the vicinity of a perfectly
conducting wall (at zero temperature), obtaining also a
transparent physical interpretation of the results in terms of
image dipoles \cite{SPR06}. In this paper we generalize this work
and consider the Casimir-Polder interaction between two
ground-state atoms at finite temperature and with boundary
conditions present, such as a conducting wall. We use a method
based on spatial correlations of the fields \cite{PT93}, which,
beside being well suited as a calculation tool for this kind of
problems, also gives a clear physical interpretation of the
results obtained. In Section \ref{Section2} we outline the method
used by reproducing in a simpler and transparent way the result
for the Casimir-Polder potential energy between two atoms in a
thermal field, well known in the literature (see, for example
\cite{WDN99}). In Section \ref{Section3} we derive and discuss our
results for the retarded atom-atom Casimir-Polder interaction when
are present both a thermal field and a boundary condition. Several
limiting cases involving the relevant parameters of the system
(temperature, atom-atom and atoms-wall distances) are explicitly
discussed.

\section{The Casimir-Polder potential between two atoms
at nonzero temperature in the free space} \label{Section2}

We first consider two neutral atoms interacting with the quantum
electromagnetic radiation field in a thermal bath at temperature
$T$, and we investigate their Casimir-Polder interaction. Our
approach to this problem exploits the idea that field fluctuations
induce instantaneous dipole moments in the two atoms, which are
correlated because vacuum fluctuations are spatially correlated.
The Casimir-Polder potential energy then arises from the classical
interaction between the oscillating dipoles of the atoms
\cite{PT93}. This method has been used in several different
contexts, such as three-body forces \cite{CP97} and time dependent
situations \cite{PPR06,RPP07}. It has been recently used also in
the case in which boundary conditions are present
\cite{SPR06,SDM07}. In this Section we show that this method is
valid and computationally useful also for the calculation of
Casimir-Polder forces at a nonzero temperature.

The relation between the Fourier components of the fluctuating
electromagnetic field (zero-point and/or thermal fields) and of
the induced dipole moment in the atoms is \cite{PT93}
\begin{equation}
\mu_l(\bk j) = \alpha (k) E_l(\bk j,\br) \label{eq:2.1}
\end{equation}
where
\begin{equation}
\alpha(k)=\frac2{3\hbar c} \sum_{p}\frac{k_{p0}\mid \mu_{p0}
\mid^2 }{k^2_{p0}-k^2} \label{eq:2.2}
\end{equation}
is the dynamical polarizability of the atoms (assumed isotropic
for simplicity), $\hbar ck_{p0}=E_p-E_0$ is the transition energy
from the state $p$ to the ground state $0$ of the atom and
$\mu_{p0}$ are matrix elements of the atomic dipole momentum
operator. $E_l(\bk j,\br)$ is the $l$ component of the electric
field operator, that in the multipolar coupling scheme coincides
with the transverse displacement field \cite{CPP95},
\begin{equation}
\mathbf E(\bk j,\br)=i\sqrt{\frac{2\pi\hbar ck}V}\ekj \left( \akj
e^{i\bk \cdot \br}-\akjd e^{-i\bk \cdot \br} \right)
\label{eq:2.3}
\end{equation}

The Casimir-Polder interaction energy, described as the classical
interaction between the induced atomic dipole moments \cite{PT93},
is then written as
\begin{equation}
\begin{split}
W_{AB}(R)&=\sum_{\bk j}\langle\mu_l^A(\bk j)\mu_m^B(\bk j) \rangle
V_{lm}(R)\\
&=\sum_{\bk j}\alpha_A(k) \alpha_B(k) \langle E_l(\bk j,\br_A)
E_m(\bk j,\br_B)\rangle V_{lm}(k,R)\end{split} \label{eq:2.4}
\end{equation}
where $R=\mid \br_B-\br_A\mid$ is the distance between the two
atoms and
\begin{equation} \begin{split}
V_{lm}(k,R)&=\FlmR \frac{\cos kR }{R} \\
&= \frac 1{R^3}\left\{ \left( \delta_{lm}-3\hat{R}_l \hat{R}_m
\right) \left( \cos kR + kR\sin kR\right) - \left(
\delta_{lm}-\hat{R}_l \hat{R}_m \right) k^2 R^2 \cos kR \right\}
\end{split} \label{eq:2.5}
\end{equation}
is the classical electromagnetic potential tensor between two
oscillating dipoles at frequency $ck$ \cite{BW80}, and the
superscript $R$ indicates the variable with respect the
derivatives are taken.

The expectation value of the spatial field correlation $\langle
E_l(\bk j,\br_A) E_m(\bk j,\br_B)\rangle$ in \eqref{eq:2.4} must
be taken on the field state in consideration, in our case the
equilibrium thermal state at temperature $T$ (isotropic and
unpolarized). Thus
\begin{equation}
\langle \akjd \akj \rangle = \frac 1{e^{\hbar ck/k_BT}-1}
\label{eq:2.6}
\end{equation}
where $k_B$ is the Boltzmann constant. We assume that the
temperature is such that the atomic excitation probability due to
the thermal field is negligible (that is, $k_BT \ll \hbar
\omega_0$, $\omega_0$ being a typical transition frequency of the
atom).

In the continuum limit, we can easily perform the polarization sum
and the angular integration
\begin{equation}
\sum_j\int d\Omega_k \langle E_l(\bk j,\br_A) E_m(\bk
j,\br_B)\rangle =\frac{8\pi^2\hbar ck}V \coth\left(\frac{\hbar
ck}{2k_BT}\right)\tau_{lm}(k,R) \label{eq:2.7}
\end{equation}
where we have used \eqref{eq:2.6} and defined the tensor
\begin{equation} \begin{split}
\tau_{lm}(k,R)&= \frac 1{4\pi}\int d\Omega_k \left( \delta_{lm}
-\hat{k}_l \hat{k}_m \right) e^{\pm i\bk \cdot
\bR}=-\FlmR\frac{\sin
kR}{k^3R} \\
&= \left( \delta_{lm}-\hat{R}_l \hat{R}_m \right) \frac {\sin
kR}{kR} +\left( \delta_{lm}-3\hat{R}_l \hat{R}_m \right)
\left(\frac {\cos kR}{k^2R^2}-\frac {\sin kR}{k^3R^3} \right)
\end{split} \label{eq:2.8}
\end{equation}

The final expression for Casimir-Polder energy, valid at any
distance $R$ between the atoms outside regions of wavefunctions
overlapping, is
\begin{equation}
\begin{split}W_{AB}(R)&=\frac{\hbar
c}\pi\int_0^\infty k^3\alpha_A(k)\alpha_B(k)\coth\left(\frac{\hbar
ck}{2k_BT}\right)V_{lm}(k,R)\tau_{lm}(k,R) \\
&= -\frac{\hbar c}\pi \int dk
\alpha_A(k)\alpha_B(k)\coth \left(\frac{\hbar ck}{2k_BT}\right) \\
&\times \left( \left( \nabla^2 \delta_{lm} -\nabla_l \nabla_m
\right)^R \frac {\cos kR}R \right) \left( \left( \nabla^2
\delta_{lm} -\nabla_l \nabla_m \right)^R \frac {\sin kR}R \right) \\
&=-\frac{\hbar c}{\pi R^3}
\int_0^\infty dk k^3\alpha_A(k)\alpha_B(k)\coth \left(\frac{\hbar
ck}{2k_BT}\right)\Big( kR \sin 2kR \\
&+2\cos 2kR-5\frac{\sin 2kR }{kR}-6\frac{\cos
2kR}{k^2R^2}+3\frac{\sin 2kR}{k^3R^3}\Big).\end{split}
\label{eq:2.9}
\end{equation}

In the so-called {\it near zone}, that is for interatomic
distances smaller than typical atomic transition wavelenghts from
the ground state, we can approximate $kR \ll 1$, and this
expression reduces to
\begin{equation}
W_{AB}(R)\simeq-\frac{3\hbar c}{\pi R^6 }\int dk
\alpha_A(k)\alpha_B(k)\coth \left(\frac{\hbar
ck}{2k_BT}\right)\sin 2kR \label{eq:2.10}
\end{equation}
which coincides with the result obtained in \cite{Milonni-Smith}
with different methods. For larger distances, in the so-called
{\it far zone}, we can replace the dynamical polarizabilities
$\alpha_{A,B}(k)$ with their static values
$\alpha_{A,B}=\alpha_{A,B}(0)$. After integration over $k$, we
obtain
\begin{equation}
W_{AB}(R)=\alpha_A \alpha_B k_BT Q^R\coth\left(\frac{2\pi
k_BTR}{\hbar c}\right) \label{eq:2.11}
\end{equation}
where the differential operator
\begin{equation}
Q^R =\left( -\frac{1}{16 r^2}\frac{\partial^4}{\partial
r^4}+\frac{1}{4 r^3}\frac{\partial^3}{\partial r^3}-\frac{5}{4
r^4}\frac{\partial^2}{\partial r^2}+\frac{3}{
r^5}\frac{\partial}{\partial r}-\frac{3}{r^6} \right)
\label{eq:2.12}
\end{equation}
has been defined. The result \eqref{eq:2.11} was already obtained
by Boyer in the framework of stochastic electrodynamics
\cite{Boyer}. Our method has allowed us to reproduce known results
in a simpler way, also obtaining a transparent physical
interpretation of Casimir-Polder forces at finite temperature in
terms of the atomic dipole moments induced by both vacuum and
thermal field fluctuations.

We can consider two limiting cases of \eqref{eq:2.11} (far zone),
given by $2\pi k_BTR/\hbar c \ll 1$ and $2\pi k_BTR/\hbar c \gg
1$; they can be considered as a low- and high-temperature limit of
the Casimir-Polder energy, respectively. Alternatively, for a
given value of the temperature, as it is known, there is a new
distance scale inside the far zone given by $\hbar c/2\pi k_BT$:
for distances smaller than this scale, equation \eqref{eq:2.11}
gives a potential energy as $R^{-7}$, while for larger distances
the potential behaves as $R^{-6}$, as in the near zone
\cite{GW99,WDN99,Barton01}.

\section{The Casimir-Polder interaction between two atoms at nonzero
temperature in the presence of a conducting wall} \label{Section3}

We now consider how the presence of a boundary condition such as a
perfectly conducting wall modifies the Casimir-Polder interaction
between the two atoms at finite temperature. We use the same
method discussed in Section \ref{Section2}. The electric field
operator is now
\begin{equation}
\mathbf E(\br ) = \sum_{\bk j} \mathbf E(\bk j,\br )=i\sum_{\bk j}
\sqrt{\frac{2\pi\hbar ck}{V}}\mathbf f(\bk j,\br) \left(\akj-\akjd
\right),\label{eq:3.1}
\end{equation}
where $\mathbf f(\bk j,\br)$ are appropriate mode functions given
by the boundary conditions for the field operators ($j$ is a
polarization index). As shown in \cite{SPR06}, the presence of the
wall requires also a modification of the classical interaction
between the induced atomic dipoles to be used in our method,
because the image dipoles (reflected on the wall) must be taken
into account. We assume the wall located at $z=0$, and that
$\br_A,\br_B$ are respectively the positions of atoms A and B.

Thus we write the atom-atom Casimir-Polder interaction energy as

\begin{equation}\begin{split}
W_{AB}(R,\Rbar )&=\sum_{\bk j}\mu_l^A(\bk j)\mu_m^B(\bk j)
V_{lm}(k,R,\Rbar )\\
&=\sum_{\bk j}\alpha_A(k)\alpha_B(k) \langle E_l(\bk
j,\br_A)E_m(\bk j,\br_B)\rangle V_{lm}(k,R,\Rbar )
\end{split}\label{eq:3.2}
\end{equation}
where the quantum average of the field operators is taken on a
thermal state of the radiation field at temperature $T$. As
already mentioned, the potential tensor $V_{lm}(k,R,\Rbar )$
should now take into account not only the interaction between the
two induced atomic dipoles, but also the interaction between the
induced dipole of one atom and the image reflected on the wall of
the induced dipole of the other atom. So we take the following
expression for the potential tensor
\begin{equation}
\begin{split}
V_{lm}(k,R,\Rbar )&=V_{lm}(k,R)-\sigma_{lp}V_{pm}(k,\Rbar ) \\
&=\left( \nabla^2 \delta_{lm} -\nabla_l \nabla_m \right)^R \,
\frac{\cos kR }R -\sigma_{lp}\left( \nabla^2 \delta_{pm} -\nabla_p
\nabla_m \right)^\Rbar \, \frac{\cos k\Rbar }\Rbar \label{eq:3.3}
\end{split}
\end{equation}
where the matrix
\begin{equation}
\sigma=\left(\begin{array}{ccc}1&0&0\\0&1&0\\0&0&-1\end{array}\right)
\label{eq:3.4}
\end{equation}
gives a reflection on the conducting plate, supposed orthogonal to
the z axis, and $\Rbar=\mid\br_B-\sigma \br_A \mid$ is the
distance between one atom and the image of the other atom
reflected on the plate. The atom-atom Casimir-Polder potential
energy \eqref{eq:3.2} adds to the well-known atom-wall
Casimir-Polder interaction energy, of course.

The equal-time spatial correlation of the electric field at points
$\mathbf{r}_A$ and $\mathbf{r}_B$ in \eqref{eq:3.2}, in the
presence of the conducting plate and evaluated on the thermal
state of the field, is
\begin{equation}
\begin{split}\langle
E_l(\bk j,\br_A )E_m(\bk j,\br_B )\rangle &=\frac{2\pi\hbar ck}V
f_l(\bk j;\br_A )f_m(\bk j;\br_B )\left( 2\langle \akjd \akj
\rangle +1 \right) \\
&=\frac{2\pi\hbar ck}V f_l(\bk j,\br_A)f_m(\bk j,\br_B)
\coth\left(\frac{\hbar ck}{2k_BT}\right)\end{split}\label{eq:3.5}
\end{equation}
In the continuum limit, performing the sum over polarizations and
the angular integration over the directions of $\bk$ and using the
appropriate mode functions \cite{PT82}, we obtain
\begin{equation}
\frac 1{4\pi}\int d\Omega_k\sum_j f_l(\bk j;\br_A )f_m(\bk j;\br_B
) =\tau_{lm}(k,R)-\sigma_{ln}\tau_{nm}(k,\Rbar)\label{eq:3.6}
\end{equation}
Substitution of \eqref{eq:3.3},\eqref{eq:3.5},\eqref{eq:3.6} into
\eqref{eq:3.2} yields
\begin{equation} \begin{split} W_{AB}(R,\Rbar)&=\frac{\hbar c}\pi\int dk
k^3\alpha_A(k)\alpha_B(k)\coth\left(\frac{\hbar ck}{2k_BT}\right)
\\
&\times \left( V_{lm}(k,R)-\sigma_{lp}V_{pm}(k,\Rbar)\right)
\left(\tau_{lm}(k,R)-\sigma_{ln}\tau_{nm}(k,\Rbar)\right)
\end{split} \label{eq:3.7}
\end{equation}
Using \eqref{eq:2.5} and \eqref{eq:2.8}, this expression can be
written in the following form
\begin{equation}
\begin{split}
W_{AB}(R,\Rbar)=&-\frac{\hbar c}\pi \int dk
\alpha_A(k)\alpha_B(k)\coth \left(\frac{\hbar ck}{2k_BT}\right) \\
&\times \left( \left( \nabla^2 \delta_{lm} -\nabla_l \nabla_m
\right)^R \frac {\cos kR}R \right) \left( \left( \nabla^2
\delta_{lm} -\nabla_l \nabla_m \right)^R \frac {\sin kR}R \right)
\\
&-\frac{\hbar c}\pi \int dk
\alpha_A(k)\alpha_B(k)\coth \left(\frac{\hbar ck}{2k_BT}\right) \\
&\times \left( \left( \nabla^2 \delta_{lm} -\nabla_l \nabla_m
\right)^\Rbar \frac {\cos k\Rbar}\Rbar \right) \left( \left(
\nabla^2 \delta_{lm} -\nabla_l \nabla_m \right)^\Rbar \frac {\sin
k\Rbar}\Rbar \right) \\
&+\frac{\hbar c}\pi\sigma_{ln}\left( \nabla^2 \delta_{lm}
-\nabla_l \nabla_m \right)^R \frac 1R \left( \nabla^2 \delta_{nm}
-\nabla_n \nabla_m \right)^\Rbar \frac 1\Rbar \\
&\times \int dk \alpha_A(k)\alpha_B(k)\sin k(R+\Rbar )\coth \left(
\frac{\hbar ck}{2k_BT} \right)
\end{split} \label{eq:3.8}
\end{equation}

Comparing \eqref{eq:3.8} with \eqref{eq:2.9}, it is evident that,
in the presence of the conducting plate, the atom-atom
Casimir-Polder potential energy at nonzero temperature is the sum
of three contributions: i) the ``direct'' interaction between the
two atoms, as in absence of the wall: this is the first term in
\eqref{eq:3.8}, which depends only on $R$ ; ii) the interaction
between an atom and the image of the other atom, which has the
same formal expression of the previous contribution, but in terms
of $\Rbar$; iii) a term depending from both distances $R$ and
$\Rbar$. In the limit of zero temperature, equation \eqref{eq:3.8}
reduces to previous results at zero temperature in reference
\cite{SPR06,PT82}.

In the far zone, the expression \eqref{eq:3.8} of the potential
energy can be written in a more compact form using the operator
defined in equation \eqref{eq:2.12}
\begin{equation} \begin{split}
W_{AB}(R,\Rbar) &= \alpha_A \alpha_B k_B T \left\{ Q^R \coth
\left( \frac R{\lambda_T} \right) +
Q^\Rbar \coth \left( \frac \Rbar {\lambda_T} \right) \right. \\
&+ \left. \sigma_{ln}\left( \nabla^2 \delta_{lm} -\nabla_l
\nabla_m \right)^R \frac 1R \left( \nabla^2 \delta_{nm} -\nabla_n
\nabla_m \right)^\Rbar \frac 1\Rbar \coth \left(\frac
{R+\Rbar}{2\lambda_T}\right) \right\}
\end{split}
\label{eq:3.9}
\end{equation}
where $\alpha_{A,B}$ are the static polarizabilities of the atoms
and we have introduced the thermal length $\lambda_T=\hbar c/2\pi
k_BT$.

It is worth to consider different limiting cases of \eqref{eq:3.9}
according to the values of $R$, $\Rbar$ and $\lambda_T$.

If $R, \Rbar \ll \lambda_T$, we obtain
\begin{equation}
\begin{split}
W_{AB}(R,\Rbar) &= -\alpha_A \alpha_B \Bigg( \frac
{23}{4\pi} \frac {\hbar c}{R^7} + \frac {23}{4\pi} \frac {\hbar c}{\Rbar^7} \\
&- \frac {\hbar c}\pi \sigma_{ln}\left( \nabla^2 \delta_{lm}
-\nabla_l \nabla_m \right)^R \left( \nabla^2 \delta_{nm} -\nabla_n
\nabla_m \right)^\Rbar \frac 1{R\Rbar (R+\Rbar)} \Bigg)
\end{split}
\label{eq:3.11}
\end{equation}
which shows that the potential in this case scales as the inverse
of the seventh power of the distance. Equation \eqref{eq:3.11}
indeed reproduces the zero-temperature result \cite{SPR06,PT82}.

If $R \ll \lambda_T$ and $\Rbar \gg \lambda_T$, equation
\eqref{eq:3.9} yields
\begin{equation}
W_{AB}(R,\Rbar) = -\alpha_A \alpha_B \left( \frac {23}{4\pi} \frac
{\hbar c}{R^7} +\frac {3k_BT}{\Rbar^6} -\frac {k_BT}{R^3\Rbar^3}
\left( 3\sin^2 \theta + 3\sin^2 \bar{\theta} -2 \right) \right)
\label{eq:3.12}
\end{equation}
where $\theta$ and $\bar{\theta}$ are respectively the angles that
$R$ and $\Rbar$ make with the axis perpendicular to the wall.
Being $R \ll \Rbar$ and $R \ll \lambda_T$, the last two terms
inside the brackets are negligible compared to the first one, and
thus the Casimir-Polder potential between the two atoms is
essentially the same as for atoms in the free space at zero
temperature.

If $R,\Rbar \gg \lambda_T$, equation \eqref{eq:3.9} yields
\begin{equation}
W_{AB}(R,\Rbar) = -\alpha_A \alpha_B k_B T \left( \frac 3{R^6}
+\frac 3{\Rbar^6} -\frac 1{R^3\Rbar^3} \left( 3\sin^2 \theta +
3\sin^2 \bar{\theta} -2 \right) \right)
\label{eq:3.10}
\end{equation}
Equation \eqref{eq:3.10} shows that in this case the potential
scales as the inverse of distance to the sixth power. We also
notice from \eqref{eq:3.10} that the last term, containing both
distances $R$ and $\Rbar$, gives a contribution to the potential
opposite to the other two terms; however, by taking into account
that $\Rbar > R$, it is easy to show that the potential is
attractive for any spatial configuration of the atoms with respect
to the wall. However, in this case both the presence of the wall
and the finite temperature of the field significantly affect the
Casimir-Polder potential energy between the two atoms.

\section{Conclusion}

In this paper we have considered the Casimir-Polder potential
energy between two atoms near a perfectly conducting plate and at
nonzero temperature, both in the near and far zone. We have
investigated the effect of the boundary condition and of the
finite temperature on the potential, in order to consider
situations close to realistic experimental setups. We have used a
method based on spatial field correlations, which has proved quite
convenient and physically transparent in dealing with this kind of
problems. Using this method we have first reproduced in a more
transparent way the known results for the atom-atom Casimir-Polder
potential in the free space at finite temperature. Then we have
applied the same method, with appropriate modifications, to derive
the expression of the atom-atom potential at nonzero temperature,
when a conducting plate is also present. We have obtained an
analytical expression of the potential both in the near and the
far zone. We have then analyzed limiting cases of interest,
according to the relation between atom-atom and atoms-wall
distances with the thermal length, which is proportional to
$T^{-1}$. In the future, we plan to extend this work to the case
in which one or both atoms are in their excited state and/or when
they are in the space between two parallel walls, where resonance
effects could yield relevant modifications of the Casimir-Polder
interatomic potential.

\begin{acknowledgments}
Partial support by Ministero dell'Universit\`{a} e della Ricerca
Scientifica e Tecnologica and by Comitato Regionale di Ricerche
Nucleari e di Struttura della Materia is acknowledged.
\end{acknowledgments}

\end{document}